\documentclass{aa} 

\usepackage[varg]{txfonts}
\usepackage{graphicx}
\usepackage{booktabs}
\usepackage{adjustbox}

\usepackage{natbib}
\usepackage{hyperref}

\newcommand{\mam}{\texttt{MAMPOSSt}}

\begin{document}

\title{Fossil group origins XIV: The radial orbits of A267}

\author{Stefano Zarattini\inst{1}, Andrea Biviano\inst{2,3}, Iacopo Bartalucci\inst{4}, J. Alfonso L. Aguerri\inst{5,6}, C. P. Haines\inst{7}, \and M. Girardi\inst{2,8}}

\institute{Centro de Estudios de F\'isica del Cosmos de Arag\'on (CEFCA), Plaza San Juan 1, 44001 Teruel, Spain
\and INAF-Osservatorio Astronomico di Trieste, via G.B. Tiepolo 11, 34143 Trieste, Italy
\and IFPU-Institute for Fundamental Physics of the Universe, via Beirut 2, 34014 Trieste, Italy
\and INAF - IASF Milano, Via A. Corti 12, 20133 Milano, Italy
\and Instituto de Astrof\'isica de Canarias, calle V\'i­a L\'actea s/n, E-38205 La Laguna, Tenerife, Spain
\and Departamento de Astrof\'isica, Universidad de La Laguna, Avenida Astrof\'isico Francisco S\'anchez s/n, E-38206 La Laguna, Spain
\and Instituto de Astronom\'ia y Ciencias Planetarias, Universidad de Atacama, Copayapu 485, Copiap\'o, Chile
\and Dipartimento di Fisica dell'Universit\'a degli Studi di Trieste Sezione di Astronomia, via Tiepolo 11, I-34143 Trieste, Italy.\\
szarattini@cefca.es}

\date{\today}

\abstract{Fossil groups (FGs) are groups or clusters of galaxies with a single, massive, central galaxy dominating their luminosity distribution, and with a clear lack of $L^*$ galaxies. The physical reason for the large magnitude gap ($\Delta m_{12}$) in these systems is still a matter for investigation. It could originate in an early formation of FGs, followed by passive evolution in which all $L^*$ galaxies merged with the central one, and/or it could be related to the fact that galaxies accreting on the FGs move on very radial orbits, reach small pericentric radii, and are merged on shorter timescales than regular cluster galaxies. The latter properties could be linked with the peculiar position of FGs within the cosmic web.}
{To shed light on the origin of FGs, we determine the velocity anisotropy profile $\beta(r)$ of the fossil cluster A267, which is related to the orbital distribution of cluster galaxies. This is the first individual FG for which the orbital distribution of its galaxies is determined. We aim to confirm previous findings based on stack samples that indicate that FGs, on average, host galaxies on more radial orbits than normal clusters.}
{We started with a sample of 2315 redshifts for galaxies in the field of A267 and we determined the membership for 329 of them. Of these, 174 are located within the virial radius of the cluster, and we used them as tracers of the gravitational potential of the cluster to solve the Jeans equation for dynamical equilibrium using the \mam\ algorithm. As a result, we obtained the cluster mass profile $M(r)$ and $\beta(r)$. We also estimated $M(r)$ from the X-ray data by applying the hydrostatic equilibrium.}
{A comparison of the \mam\ and X-ray-determined $M(r)$s allows us to estimate the cluster hydrostatic mass bias, which we find to be consistent with previous findings. The anisotropy parameter $\beta(r)$ indicates tangential orbits for the galaxies near the cluster centre and increasingly radial orbits in the external regions. We checked that our results are not affected by the presence of subclusters and by the choice of the models for $M(r)$ and $\beta(r)$.}
{The A267 $\beta(r)$ is very similar to that previously determined for a stack of large $\Delta m_{12}$ systems. Our analysis therefore confirms that FGs are characterized by more radial orbits for their member galaxies than the average cluster population. We speculate that this different orbital distribution might be an important element in creating a large $\Delta m_{12}$.
}

\keywords{}

\authorrunning{S. Zarattini, et al.}
\titlerunning{Radial orbits in A267}
\maketitle

\section{Introduction}
\label{sec:intro}
Galaxy clusters are the most massive virialized objects in the Universe. They are defined by a deep potential well, which is able to confine, within a relatively small scale of $\sim 1$ Mpc radius, hot intracluster gas \citep[which accounts for $\sim 15\%$ of the total cluster mass, see e.g. ][]{Allen2011}, galaxies \citep[up to several thousand, but accounting for only $\sim 1-3\%$ of the mass, see e.g. ][]{Lin2004}, and dark matter (the remaining -- and dominant -- fraction of the mass).

It is well known that galaxies undergo a series of physical processes in clusters that strongly impact many of their characteristics. For example, differences in the luminosity \citep[e.g.][]{Adami1998a}, colour \citep[e.g.][]{Blanton2005}, morphology \citep[e.g.][]{Dressler1980}, star formation activity \citep[e.g.][]{Hashimoto1998}, and gas content \citep[e.g.][]{Giovanelli1985} are directly impacted by the cluster's strong gravitational potential as well as by galaxy-galaxy interactions, which happen more frequently in clusters due to their high galaxy density.
Galaxies in clusters tend to be redder, with older stellar populations and earlier morphologies than galaxies in the field.

Within this context, studying the orbits of galaxies is a useful tool to understand their evolution mechanisms. Galaxies fall into the cluster gravitational potential 
from their position within the cosmic web.
Those that are now located in the central cluster regions arrived at an early time and had sufficient time to relax onto isotropic orbits \citep[i.e. randomly distributed on all orbital shapes,][]{Mamon2019}, whereas galaxies in the cluster outskirts were accreted more recently and still display more radially elongated orbits. However, it is not yet clear which mechanism is responsible for shaping the orbits of the accreted galaxies,  which can reach the cluster from the field, along more or less well-defined filaments, as individuals 
or as part of galaxy groups. 

Measuring the elongation of galaxy orbits is a useful tool to investigate galaxy evolution in clusters. In fact, 
various studies found that the orbits of cluster galaxies can be different according to the galaxy properties. Galaxies with different morphologies are found on different orbits; spirals and star-forming galaxies generally display more radial orbits than ellipticals and passive galaxies, respectively \citep[e.g.][]{Biviano2004,Mamon2019,ValkRembold2025}. Ram-pressure stripped galaxies are the cluster galaxy population characterized by the most elongated radial orbits \citep{Biviano2024}. According to \citet{Muriel2025},  these galaxies were accreted recently, as ram-pressure stripping is effective during the first crossing. 

\begin{figure}
 \begin{center}       
    \includegraphics[width=0.5\textwidth]{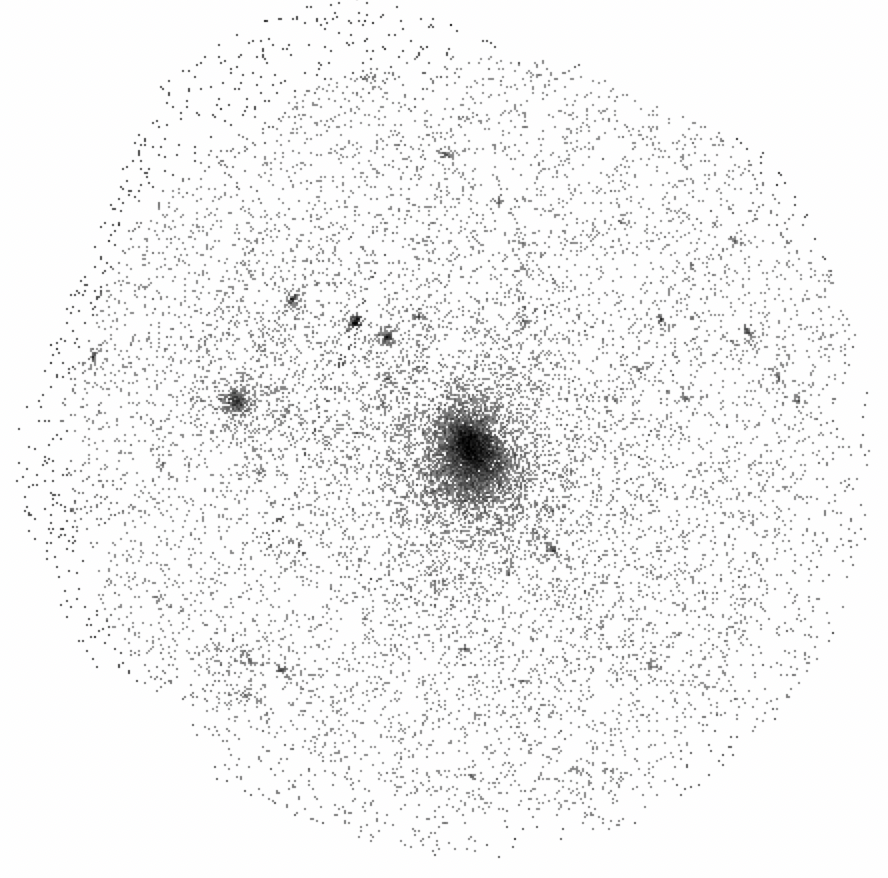}\caption{XMM-Newton exposure-corrected count-rate image in the [0.3-2] keV band.  The image was obtained by combining the MOS1, MOS2, and pn cameras.}
     \label{fig:XMM}
         \end{center}
\end{figure}

As more recently accreted galaxies are found at larger cluster-centric distances and they are characterized by radial orbits, the general cluster population displays an increasingly radial orbital distribution at larger radii \citep[e.g.][]{Biviano2013,Aguerri2017,Li2023}. However, limiting the analysis of galaxy orbits to the virial region might not show this radial dependence, since the infall region of clusters extends beyond the virial radius \citep{Rines2000,Santucho2020,Mpetha2024}. In fact, in the analysis of a stack of galaxy clusters limited to their virial region, \citet{Zarattini2021} found no significant radial dependence of galaxy orbits in clusters with a magnitude gap (the difference in magnitude, in the $r-$band, between the two brightest member galaxies within half the virial radius) of $\Delta m_{12} < 1.5$. On the other hand, \citet{Zarattini2021} did find increasingly radial orbital anisotropy with cluster-centric distance for galaxies in clusters with $\Delta m_{12} > 1.5$. 

The more radial orbits of galaxies in systems with a larger $\Delta m_{12}$ could provide an explanation for the formation of fossil groups (FGs), which are groups or clusters characterized by $\Delta m_{12} > 2$ measured on galaxies located within half the virial radius of the system centre. As suggested by \citet{D'Onghia2005} and \citet{Sommer-Larsen2005}, the large magnitude gap of FGs could be created by the tidal disruption of infalling galaxies reaching very close to the cluster centre because of their very radially elongated orbits. In this scenario, FGs should not longer be considered old and dynamically relaxed fossil relics of the ancient Universe \citep{Ponman1994}; rather, they could be galaxy groups or clusters that are more isolated from the cosmic web, which induces the infalling of galaxies on more radial orbits and leads to the creation of the magnitude gap in recent times \citep{Zarattini2023}.

The results of \citet{Zarattini2021} on the orbits of galaxies in clusters of different magnitude gaps were based on the stacking of about 100 galaxy clusters. To ensure sufficient statistics, \citet{Zarattini2021} considered four bins of $\Delta m_{12}$, including all systems with $\Delta m_{12} > 1.5$ in the largest magnitude gap bin, which therefore contain fossil as well as non-fossil systems. The aim of the present work is to confirm the results of \citet{Zarattini2021} by determining the orbits of the galaxies in a spectroscopically confirmed FG, Abell 267 \citep[][A267 hereafter]{Abell1958}. This cluster is  part of the FOssil Group Origins project \citep[FOGO, ][]{Aguerri2011}, also indicated with the name  FGS02 in other FOGO publications \citep[e.g.][]{Zarattini2014,Zarattini2016}, and it has $\Delta m_{12} > 2.12$ \citep{Zarattini2014}. In Sect. \ref{sec:members} we recompute the magnitude gap of A267 with the new data.
 
This paper is organized as follows. In Sect. \ref{sec:sample}, the optical and X-ray data are presented. In Sect. \ref{sec:method}, we present our analysis; how we identified the cluster members, the subclusters inside the clusters, and how we performed the dynamical analysis. Our results are given in Sect.~\ref{sec:results} and discussed in
Sect.~\ref{sec:discussion} where we also draw our conclusions. Throughout this paper, as in all the FOGO papers, we adopt a flat $\Lambda$CDM cosmology with $\Omega_{m} = 0.3$, $\Omega_{\Lambda} = 0.7$, and $H_{0} = 70$ km s$^{-1}$ Mpc$^{-1}$ \citep{Kauffmann2003,Aguerri2011,Paris2012}, which leads to a scale of 3.674 kpc/" at the redshift of A267 ($z=0.23$).

\section{The data}
\label{sec:sample}

In this section we present the optical and X-ray data used for this work (Sect. \ref{sec:optical} and Sect. \ref{sec:xray}, respectively). The former are used to compute the radial anisotropy profile, the latter to define the centre of the cluster and its mass profile. 

\begin{figure}
    \centering
    \includegraphics[width=0.5\textwidth, trim=180 130 180 150]{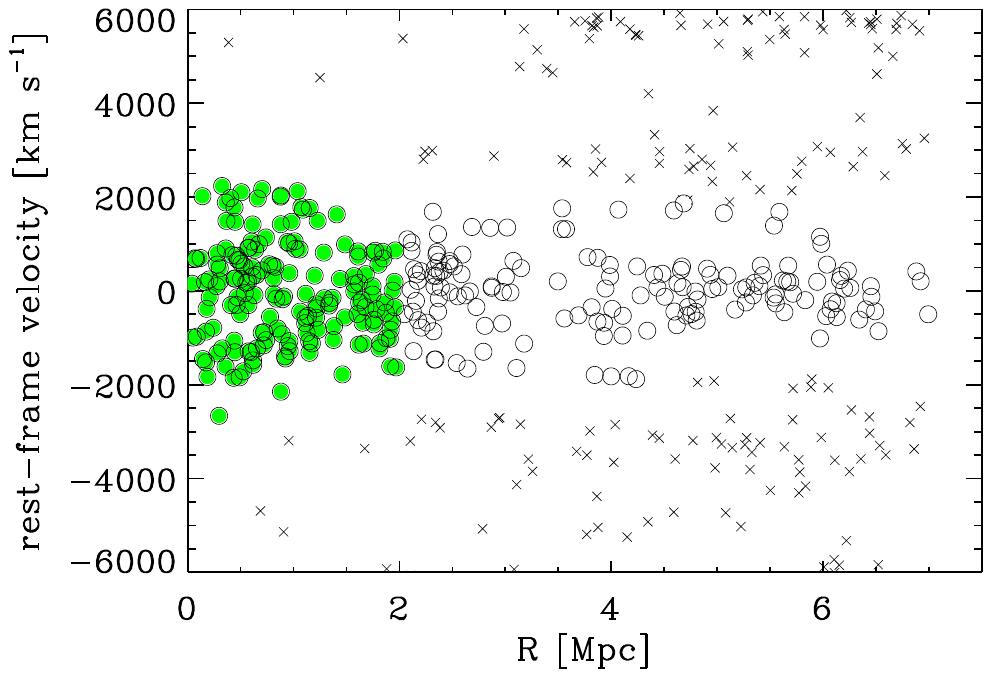}
    \caption{Distribution of galaxies in the projected phase-space of A267. Cluster members within (respectively beyond) 2 Mpc from the cluster centre are shown as green filled dots (respectively open circles). Interlopers are shown as crosses.}
    \label{fig:rvmem}
\end{figure}

\subsection{Optical data}
\label{sec:optical}

\begin{figure}
    \centering
    \includegraphics[width=0.5\textwidth, trim=200 130 150 140]{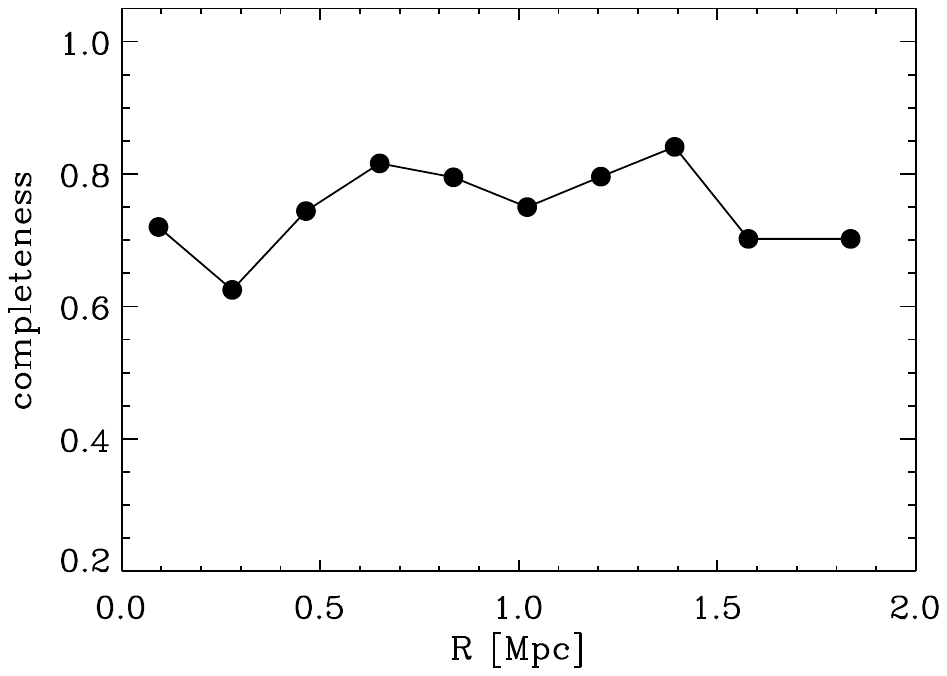}
    \caption{Completeness of the spectroscopic sample as a function of projected cluster centric distance, R. }
    \label{fig:rvmem}
\end{figure}

Our sample was built by cross matching and merging different catalogues available in the literature. The starting point was the catalogue of redshifts obtained as part of the Local Cluster Substructure Survey (LoCuSS). Three configurations were taken with the Hectospec spectrograph on the 6.5m Multi-Mirror Telescope (MMT) in Arizona in 2009, obtaining 504 redshifts. A267 was one of 30 galaxy clusters at $0.15<z<0.30$ that were observed with the Hectospec instrument as part of the Arizona Cluster Redshift Survey (ACReS). These data were used in \citet{Haines2013,Haines2015}.
Targets were selected using near-infrared photometry obtained with the NEWFIRM imager on the 4m Mayall telescope at the Kitt Peak Observatory in Arizona, as being located along the J-K versus K-band colour-magnitude relation of galaxies at the cluster redshift \citep[see][for details]{Haines2009}. The 504 redshifts in this catalogue reach a maximum distance of about 2.5 $r_{200}$\footnote{The radius $r_{\Delta}$ is the radius of a sphere with a mass overdensity $\Delta$ times the critical density at the redshift of the galaxy system.} from the centre of A267.

We then merged this first catalogue with the data obtained by \citet{Rines2013}, who used the same Hectospec instrument to get 1198 redshifts out to about 3.8 $r_{200}$. A total of 251 galaxies from the LoCuSS catalogue also have a redshift in the \citet{Rines2013} catalogue. We were thus able to add 947 redshift from the \citet{Rines2013} catalogue to the LoCuSS one, getting a total of 1451 redshifts.

Then, since A267 was also observed as part of the FOGO project, we cross matched this catalogue with the one obtained in the previous paragraph. The photometric observations were taken at the Nordic Optic Telescope (NOT) in the period 2008-2011, whereas the spectroscopic observations were taken at the Telescopio Nazionale Galileo (TNG) telescope in the period 2008-2010. In \citet{Zarattini2014}, these data were matched with the Sloan Digital Sky Survey Data Release 7 \citep[SDSS-DR7,][]{Abazajian2009} with the aim of completing the photometric and spectroscopic samples. The resulting FOGO catalogue consists of 6632 entries, with redshifts for galaxies in the magnitude range $14.5 \lesssim r \lesssim 21.5$. According to Table 1 of \citet{Zarattini2014}, there are 111 galaxies with measured redshifts within $r_{200}$ of the A267 centre, of which 42 are considered to be cluster members. The virial radius of A267, which we assume to coincide with $r_{200}$, is estimated to be 1.85 Mpc by \citet{Zarattini2014}, which corresponds to 8.4 arcmin given the cluster redshift $z_{\rm{A267}}=0.23$. These values are updated in our dynamical analysis (see Sect. \ref{sec:method}). When cross matching the FOGO catalogue with the previous one, built on the Hectospec observations, we were able to add 25 redshifts to the sample, reaching a total of 1476 redshifts.

Finally, we included the Dark Energy Spectroscopic Instrument Data Release 1 (DESI-DR1) data \citep{DESIDR1} in the area of A267. The DESI-DR1 dataset consists of more than 18 million spectra of which 13 millions are galaxies. We used data out to about 3.5 $r_{200}$, for a total of 1629 redshifts. Of those, 839 are new redshifts, not available in the catalogue built in the previous passages. It is worth noting, however, that almost all of these new redshifts are outside the virial radius of the cluster.
The final spectroscopic catalogue contains 2315\footnote{The redshift catalogue used in this work is only available in electronic form at the CDS via anonymous ftp to cdsarc.u-strasbg.fr (130.79.128.5) or via http://cdsweb.u-strasbg.fr/cgi-bin/qcat?J/A+A/.} galaxies in a circular area of radius $\sim 4 \, r_{200}$ around the cluster centre (defined by the X-ray analysis described in Sect.~\ref{sec:xray}).
\begin{figure}
    \centering
    \includegraphics[width=0.5\textwidth, trim= 190 140 180 140]{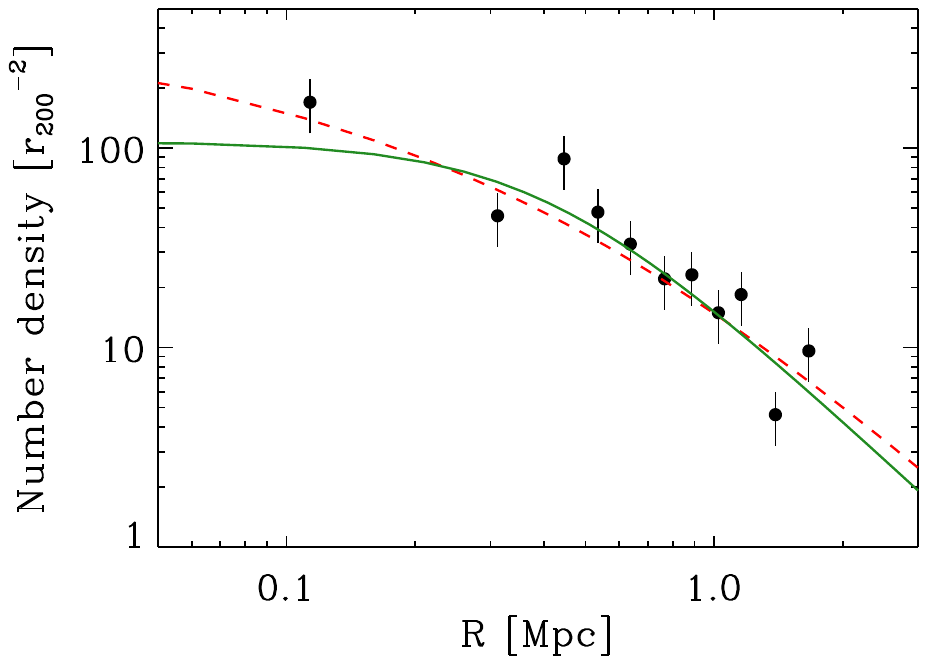}
    \caption{Projected number density profile of A267 member galaxies, corrected for incompleteness (dots with 1 $\sigma$ error bars) and best fits with a projected NFW (dashed red line) and a King model (solid green line).}
    \label{fig:NR}
\end{figure}

\subsection{X-ray data}
\label{sec:xray}
We used the archival XMM-Newton observation ID 0084230401 to derive an alternative and independent estimation of the A267 mass profiles.
The cleaning and preparation of the dataset are detailed in appendices A2, A3, and A4 of \citet{Bartalucci2017}. We combined the observations taken with MOS1, MOS2, and pn after the cleaning procedures, which yielded a useful exposure time of $\sim 26$ ks. All the results shown in this work are based on this combined dataset. We determined the X-ray peak as the peak of the emission in the exposure-corrected image shown in Fig. \ref{fig:XMM}.

We followed the techniques described in \citet{Croston2006,Croston2008} and \citet{Bartalucci2018} to derive the density and temperature radial profiles. With these profiles in hand, we computed the mass profile under the assumption of hydrostatic equilibrium following the technique detailed in \citet{Democles2010} and \citet{Bartalucci2018}.
The error bars associated with the mass profile shown in Fig. \ref{fig:mass_profile} are at 1$\sigma$ significance level.

\section{Analysis}
\label{sec:method}
\subsection{Cluster membership} 
\label{sec:members}
To identify the cluster members among the 2315 galaxies with available redshifts in the cluster field, we ran the \texttt{Clean} procedure \citep{Mamon2013} in projected phase-space (rest-frame line-of-sight velocity $v_{\rm{rf}}$ vs. projected cluster-centric distance, $R$). We identified 329 cluster members, of which 174 are located at $\leq 2$ Mpc from the cluster centre (filled green circles in Fig. \ref{fig:rvmem}). The mean cluster redshift estimated on the 329 members is 0.23007, which corresponds to a mean velocity of 68972 km~s$^{-1}$.

With the new members selection, we are now able to measure a robust magnitude gap for A267. In fact, in \citet{Zarattini2014} the authors presented a lower limit to the magnitude gap ($\Delta m_{12} > 2.12$). Now, we can confirm its fossil status and compute a new magnitude gap of $\Delta m_{12} = 2.46$.

\begin{table}
\centering
\caption{Results from the $N(R)$ fits and \mam.}
\label{tab:mamposst}
\begin{adjustbox}{width=\columnwidth,center}
\begin{tabular}{c|cc|cc}
\toprule
$N(R)$ & \multicolumn{2}{c}{$M(r)$ NFW} & \multicolumn{2}{c}{$\beta(r)$} \\
$r_{\nu}$ & $r_{200}$ & $r_s$ & $(\sigma_r/\sigma_\theta)_0$ & $(\sigma_r/\sigma_\theta)_\infty$ \\
\midrule
0.35-0.46 (King)~ & 1.90-2.08 & 0.2-0.6 & 0.5-0.8 & 1.3-9.1 \\
0.39-0.70 (NFW)   & 1.85-2.01 & 0.2-0.6 & 0.5-1.0 & 1.3-9.1 \\
\bottomrule
\end{tabular}
\end{adjustbox}
\tablefoot{We report the marginalized $\pm 1 \sigma$ intervals.}
\end{table}

\subsection{Number density profile}
\label{sec:profile}
The spectroscopic follow up of A267 was obtained using different telescopes and instruments, as discussed in \ref{sec:optical}. It is therefore crucial to estimate the spectroscopic completeness of the final sample in order to minimize the effect that incompleteness can have on the dynamical analysis.  We computed  the spectroscopic completeness as a function of the radial distance from the cluster centre (the position of the X-ray emission peak, see Sect.~\ref{sec:xray}) by combining SDSS and DESI photometry. Following the SDSS data model, the former catalogue has a $5 \sigma$ depth of $r=22.7$ mag, whereas the latter has a $5 \sigma$ depth of $r=23.4$ mag \citep{Dey2019}.

The final photometric catalogue, which was obtained by combining the two catalogues described in the previous paragraph, was built from about $30\,000$ objects. We used this catalogue to compute the radial spectroscopic completeness in annuli of 0.1 $r_{200}$, starting from the centre of the cluster. We obtained completeness values in the range of $60-84$\% for the spectroscopic sample of $m_r \leq 20.25$  galaxies within $r_{200}$ (see Fig. \ref{fig:rvmem}).

We fitted the completeness-corrected number density profile, $N(R),$ of cluster members with two models: the centrally cuspy NFW model of \citet{Navarro1996}
in projection \citep{Bartelmann1996}, 

\begin{equation}
N(x) \propto \frac{r_{\nu}}{x^2 - 1} \, f(x),
\end{equation}

where $x=R/r_{\nu}$, $R$ is the projected cluster-centric distance, and

\begin{equation}
f(x) = 
\begin{cases}
1 - \frac{2}{\sqrt{x^2 - 1}} \arctan \sqrt{\frac{x - 1}{x + 1}} & (x > 1) \\
1 - \frac{2}{\sqrt{1 - x^2}} \, \text{arctanh} \sqrt{\frac{1 - x}{1 + x}} & (x < 1) \\
0 & (x = 1),
\end{cases}
\end{equation}

and the  cored King model \citep{King1962}

\begin{equation}
N(x) \propto (1+x^2)^{-1}.
\end{equation}

The fits were performed without binning the data, following the maximum likelihood procedure of \citet{Sarazin1980}. The normalization of the profile is not a free parameter, as it is set by the requirement that the integral of the density profile corresponds to the total, completeness-corrected number of galaxies. In any case, the normalization of the number density profile does not matter in the dynamical analysis (Sect.~\ref{sec:MAMPOSSt}), since the number density profile only enters the Jeans equation through the logarithmic derivative \citep{Binney1987}. So the only free parameter in the fits is the scale radius of the galaxy number density profile, $r_{\nu}$.
The quality of the fits is evaluated by a $\chi^2$ on radially binned profiles, which is illustrated in Fig.~\ref{fig:NR}. Both the NFW and the King models provide acceptable fits to the data, even if the King model is marginally better than the NFW one, with a $\chi^2=14.8$ versus 14.9, for 11 degrees of freedom. We therefore perform our dynamical analysis by considering both the best-fit King and the best-fit NFW model solutions
(see Table~\ref{tab:mamposst}).

\subsection{MAMPOSSt}
\label{sec:MAMPOSSt}

To estimate the orbits of galaxies in A267, we solved the Jeans equation for dynamical equilibrium in spherical symmetry \citep{Binney1987}. The spherical assumption is an approximation, since the X-ray image shows an elongation in the NE-SW direction. However,
our dataset of cluster members does not provide sufficient statistics to go beyond the spherical approximation. As regards the hypothesis of dynamical equilibrium, in Sect. \ref{sec:discussion} we are going to show  that removing optically identified subclusters from our analysis does not have a significant effect on the results of our dynamical analysis. 

We used \mam\ \citep{Mamon2013} to solve the Jeans equation. Given
models for both $M(r)$ and $\beta(r)$, \mam\ estimates the joint probability of finding the member galaxies at their observed positions in the projected phase-space diagram. To assess the quality of the results, \mam\ was tested on simulated haloes, which provided robust results for both relaxed and unrelaxed systems \citep{Mamon2013,Aguirre2021}. Moreover, it was also applied to several observational datasets and it is now a consolidated method for determining the orbital shape of galaxies in clusters \citep{Biviano2013,Guennou2014,Munari2014,Verdugo2016,Biviano2017,Mamon2019,Sartoris2020}.

To define the positions of member galaxies in projected-phase space that constitute the observables on which \mam\ operates,  we needed to adopt a cluster centre in right ascension, declination, and velocity. We fixed the cluster centre in right ascension and declination to the peak of the X-ray emission (see Sect.~\ref{sec:xray}) and the velocity centre to the mean velocity of cluster members (see Sect.~\ref{sec:members}). Since the Jeans equation is valid in conditions of dynamical equilibrium, we avoided including external cluster regions in the \mam\ analysis and only considered galaxies with $R \leq 2$ Mpc, of which there are 174.

For $M(r)$ we adopted the NFW model, which is a very good description of the distribution of dark matter in simulated cosmological haloes \citep{Navarro1997} and has been shown to provide adequate fits to the total mass profile of clusters of galaxies \citep[e.g.][]{Ettori+19}. The NFW model is also one of the two models we adopted to describe the spatial distribution of cluster galaxies (see Sect.~\ref{sec:profile}). Given that the total mass is dominated by the dark matter component \citep[see, e.g.,][]{BS06}, there is no reason to expect the distribution of cluster galaxies to be similar to the distribution of the total mass, so we also considered the King model to describe the spatial distribution of cluster galaxies (see Sect.~\ref{sec:profile}).

The NFW model is characterized by two free parameters, the virial radius, $r_{200}$ , and the scale radius, $r_s$.
We adopted Gaussian priors for $r_{200}$, with a mean of 1.87 Mpc and a rms of 0.15 Mpc, which are consistent with previous estimates from \citet{Rines2013} and \citet{Haines2018}.
On the other hand, there is no robust estimate of $r_s$ so we adopted generous flat priors for this parameter, 0.1 to 2.0 Mpc.

For the anisotropy profile, we used the generalised profile presented in \citet{Tiret2007}:

\begin{equation}
    \beta = \beta_0 + \beta_\infty r / (r + r_\beta)
    \label{eq:tiret}
,\end{equation}
assuming $r_\beta = r_s$ to reduce the free parameters in the \mam\ analysis, in agreement with the results from numerical simulations \citep{Mamon2010}. Rather than adopting $\beta_0$ and $\beta_\infty$ as free parameters, we prefer to use the ratio of the radial to tangential velocity dispersions, $\sigma_r/\sigma_{\theta}$ at $r=0$ and $r \rightarrow \infty$, that is $(\sigma_r/\sigma_\theta)_0$ and $(\sigma_r/\sigma_\theta)_{\infty}$ for which we adopt flat priors in the ranges 0.3-2.0 and 0.6-15.0, respectively.
We ran a Monte Carlo Markov chain (MCMC) procedure with 100,000 steps, using the \citet{GR92} criterion to check for convergence.

\begin{figure}
 \begin{center}        
    \includegraphics[width=0.5\textwidth, trim=200 130 170 150]{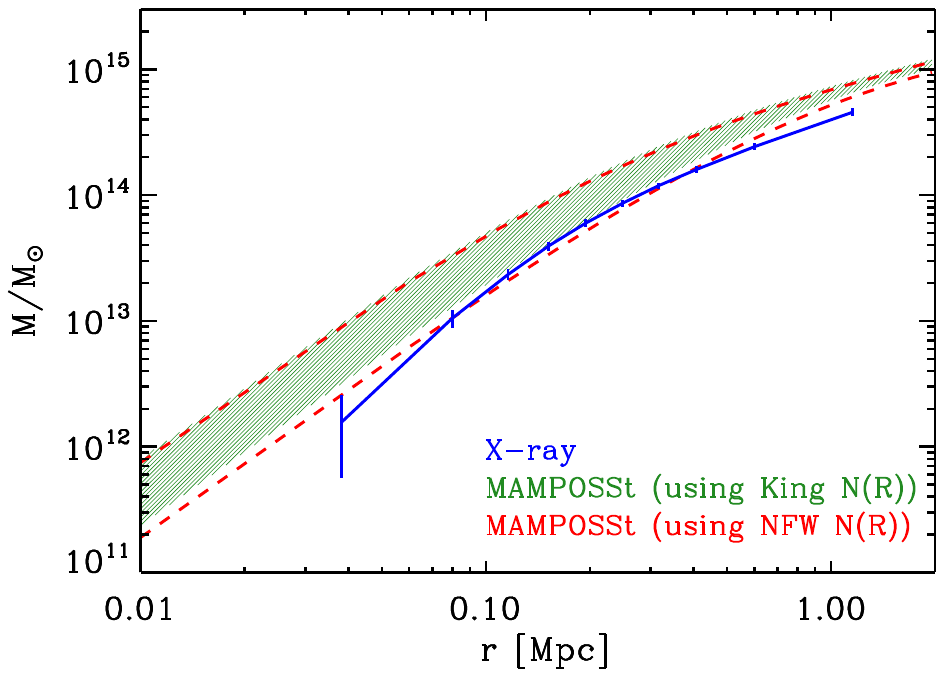}\caption{A267 $M(r)$ as derived from X-ray data (blue points with one sigma error bars; the solid blue line connects the profile points and is shown for the sake of clarity only) and from \mam, using the King and NFW models for $N(R)$ (hatched green region and dashed red lines, respectively),} indicating the $1 \sigma$
    interval derived by randomly selecting 5000 $r_{200}, r_s$ values of the NFW profile from
    the MCMC steps and taking the 16-84 percentiles of the mass values at 40 radial positions.
     \label{fig:mass_profile}
         \end{center}
         \end{figure}
         
\section{Results}
\label{sec:results}
While the focus of this work is on $\beta(r)$, \mam\ also provides us with an estimate of $M(r)$ that we can compare with the independent estimate obtained using X-ray data and assuming hydrostatic equilibrium. The \mam\ and X-ray $M(r)$ are shown in Fig. \ref{fig:mass_profile}. The \mam\ $M(r)$ derived using the King and NFW models for the spatial distribution of galaxies are almost identical.
The X-ray-derived $M(r)$ is smaller than the \mam\ one. A comparison of the two profiles allows us to measure the hydrostatic mass bias $b = 1-M_{\rm{Hyd}}/M_{\rm{MAM}}$, where $M_{Hyd}$ and $M_{\rm{MAM}}$ are the hydrostatic and  \mam\ profiles, respectively. We find that the bias varies from $b=0.5 \pm 0.3$ in the central region (0.08 Mpc) to $b=0.3 \pm 0.3$ in the intermediate regions around $r_s$, and to
$b=0.35 \pm 0.09$ in the most external regions covered by X-ray observations ($\simeq 1$ Mpc). Our values are in agreement with previous observational estimates 
\citep[see e.g. Fig.~3 in][]{Lesci2023} as well as with estimates from numerical simulations \citep[see e.g. Fig.~4 in][]{Barnes2021}, which typically find $b \approx 0.3$.

\begin{figure}
    \centering
    \includegraphics[width=0.5\textwidth, trim= 200 120 170 150]{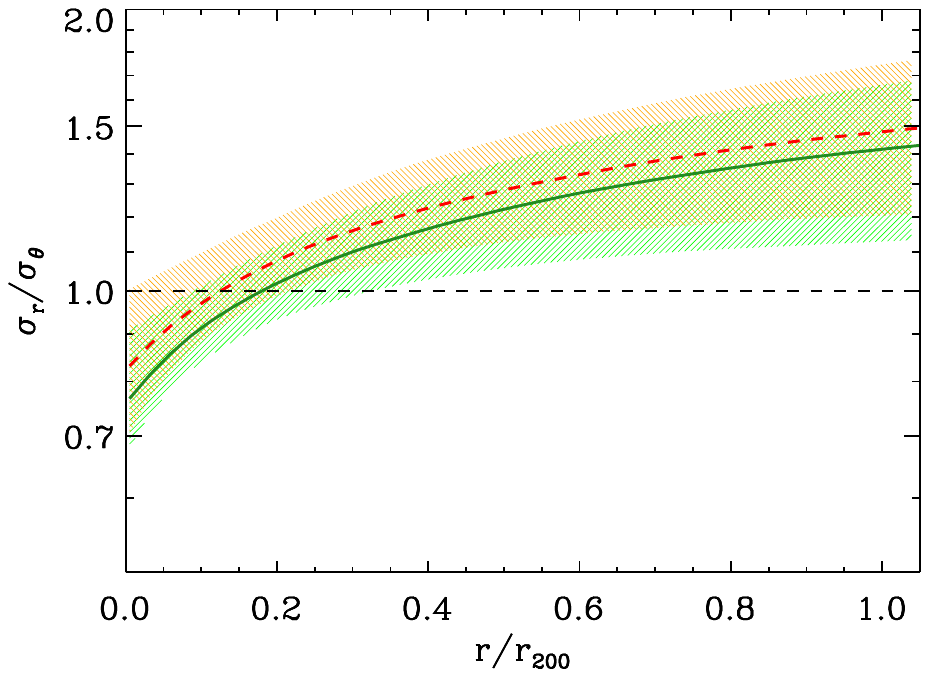}
    \caption{Velocity anisotropy profiles of A267, within 1 $\sigma$ confidence interval (shadings), derived by randomly selecting 5000 $(\sigma_r/\sigma_\theta)_0, (\sigma_r/\sigma_\theta)_\infty$ values of the generalized Tiret profile from
    the MCMC steps and taking the median and the 16-84 percentiles of the $\sigma_r/\sigma_\theta$ values at 40 radial positions. We show in green (respectively, in red) the results obtained by adopting the King (respectively the NFW) $N(R)$ model in the \mam\ analysis.}
    \label{fig:beta_profile}
\end{figure}

In Fig. \ref{fig:beta_profile} we show the radial anisotropy profile $\beta'(r)$ of A267. In this case as well, the solution does not depend in a significant way on the model adopted to describe the spatial distribution of galaxies.
Orbits near the cluster centre are preferentially tangential, ($\beta' \equiv \sigma_r/\sigma_\theta < 1$). Removing galaxies near the centre ($R<0.05$ Mpc) from the analysis reduces the tangential component but not significantly so. Orbits become more radial with increasing distance from the cluster centre, 
reaching $\sigma_r/\sigma_\theta \simeq  1.4$ at $r=r_{200}$. 

The A267 velocity anisotropy profile is in good agreement with the one we found for a stack of 
systems with $\Delta m_{12} > 1.5$  \citep{Zarattini2021}. In that paper, the authors found no radial dependence of the velocity anisotropy profile for systems with $\Delta m_{12} < 1.5$, whereas for systems with $\Delta m_{12} > 1.5$ they found 
$\sigma_r/\sigma_\theta \simeq 0.8$ near the centre and $\sigma_r/\sigma_\theta \simeq 2.0$ at $r=r_{200}$, which overlaps with the velocity anisotropy profile of A267. 
Our study of an individual fossil system with extensive spectroscopic data therefore confirms our previous finding based on a stack of several, poorly sampled FGs. It is worth noting that A267 was part of the sample of \citet{Zarattini2021}, contributing to the sample of $\Delta m_{12} > 1.5$ clusters with 45 out of 792 member galaxies.

\section{Discussion}
\label{sec:discussion}

In this section we discuss the systematics of our method as well as the consequences of our results in the context of FG formation and evolution. In particular, in Sect. \ref{sec:relax} we discuss the impact of the presence of substructures in our results. In Sect. \ref{sec:betadiffmodels} we analyse how different models for the mass and anisotropy profiles modify our results. Finally, in Sect. \ref{sec:FG_model} we discuss the framework of FG formation and evolution and how our results can be interpreted in that framework.

\subsection{Dynamical relaxation}
\label{sec:relax}
Our results are based on the \mam\ technique that relies on the Jeans equation for dynamical equilibrium. Since dynamical equilibrium is not expected beyond the virial region, in our dynamical analysis we only included galaxies with $R \lesssim r_{200}$. Here we discuss the possibility that A267 is in fact not fully dynamically relaxed even within the virial region.

The comparison of the two $M(r)$ obtained by the application of \mam\ on the phase-space distribution of cluster members and by the application of the hydrostatic equilibrium equation to the X-ray data for the intra-cluster gas, indicates a hydrostatic mass bias that is not at all anomalous, as it could be expected for a very unrelaxed cluster and is in line with previous determinations for large cluster samples (see Sect.~\ref{sec:results}). On the other hand, according to \citet{Jeltema2005}, the X-ray morphology of A267 suggests a disturbed single-component cluster, with an elongation in the NE-SW direction. The
absence of a cool core is a further argument against full dynamical relaxation \citep{Jimenez-Bailon2013}. 

The small separation between the X-ray centre and the brightest cluster galaxies (BCG), $\simeq 40$ kpc, which is similar to the average estimated in 87 eRosita Final Equatorial Depth Survey (eFEDS) clusters \citep[$76^{+30}_{-27}$ kpc, see][]{Seppi2023}, is instead considered evidence of a relaxed dynamical state according to \citet{Ota2023}. The BCG is also very close to the cluster centre in velocity space, since its rest-frame velocity difference with respect to the cluster mean velocity is 153 km~s$^{-1}$. This value is in agreement with previous studies in the literature \citep[see e.g.][who find typical BCG rest-frame velocities of $100-200$ km s$^{-1}$]{DePropris2021}.

Even if the BCG appears to be very close to the cluster centre in projected-phase space, the whole galaxy distribution of 
A267 shows significant evidence of substructures. In \citet{Zarattini2016} the authors identified a subcluster in the N direction with respect to the cluster centre using the \citet{Dressler1988} and \texttt{2D-DEDICA} \citep{Pisani1996} tests. More recently, \citet{Tucker2020} also found up to four distinct galaxy sub-populations. With the new data in hand, we ran the \texttt{DS+} method \citep{Benavides2023} to further check for the presence of subclusters in the phase-space distribution of cluster members. The \texttt{DS+} method is a development of the classical one by \citet{Dressler1988} that enables the identification of individual galaxies that are assigned to a significant subcluster by the method. 

We ran \texttt{DS+} with 1000 Monte Carlo resamplings and identified all subclusters that were deemed significant with a probability $<1/1000$. Since we only want to identify the main subclusters, we only considered possible groups of multiplicity 
$\geq 17$, that is 1/10 of the total number of members up to a maximum value of 58, which is 1/3 of the total number of members.

\begin{figure}
    \centering
    \includegraphics[width=0.5\textwidth, trim=20 20 180 0]{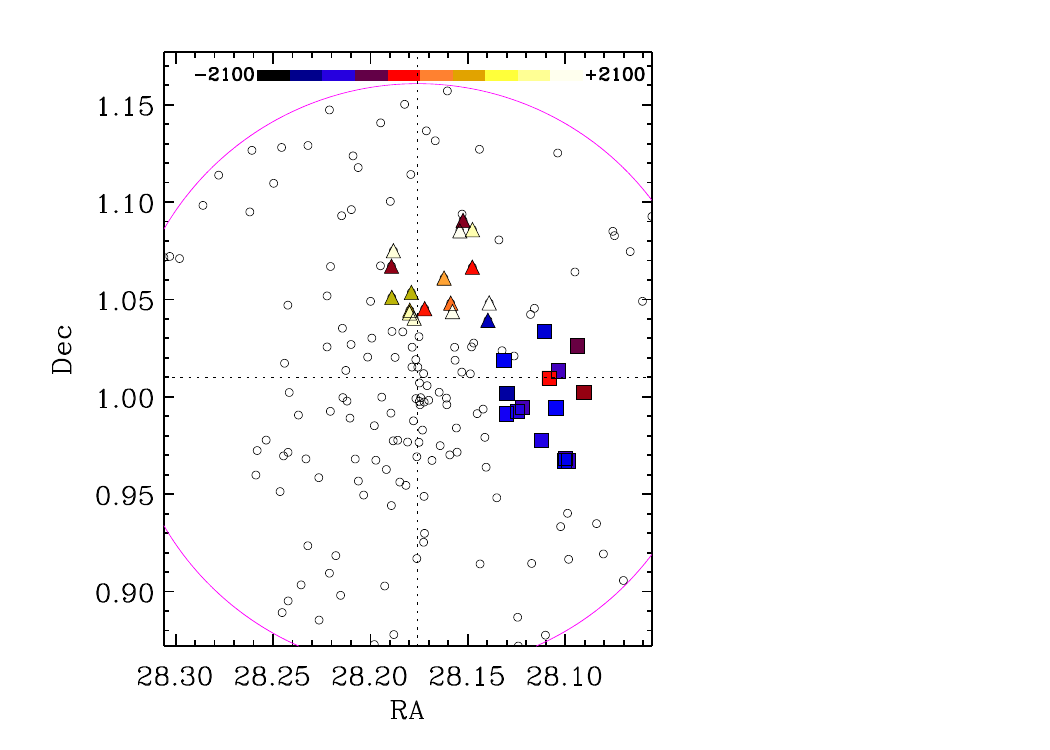}
    \caption{The result of the \texttt{DS+} test. The galaxies assigned to the two significant subclusters are indicated by coloured squares (the W group) and triangles (the N group) and colour-coded by rest-frame velocity as indicated by the coloured sidebar at the top. The violet circle represent the virial radius of A267.}
    \label{fig:sub_optical}
\end{figure}

In Fig. \ref{fig:sub_optical} we present the results of our \texttt{DS+} analysis. The method identifies two significant subclusters: one, the most significant, to the west of the cluster centre (hereafter the W group), and the other to the north-east of the cluster centre (hereafter the N group). The W group has a mean velocity of $-1193 \pm 70$ km s$^{-1}$ and a velocity dispersion of $260_{-44}^{+52}$ km s$^{-1}$ in the cluster rest frame. The N group has a mean rest-frame velocity of $+875 \pm 300$ km s$^{-1}$ and a velocity dispersion of $1191_{-190}^{+226}$ km s$^{-1}$. The kinematics of the two groups suggest that the W one is infalling into the cluster from the foreground for the first time, while the N one has already passed the cluster centre, the pericentric passage has disrupted the compactness of its velocity distribution, and it is now moving out. 

\begin{figure}
    \centering
    \includegraphics[width=0.5\textwidth, trim= 30 20 15 0]{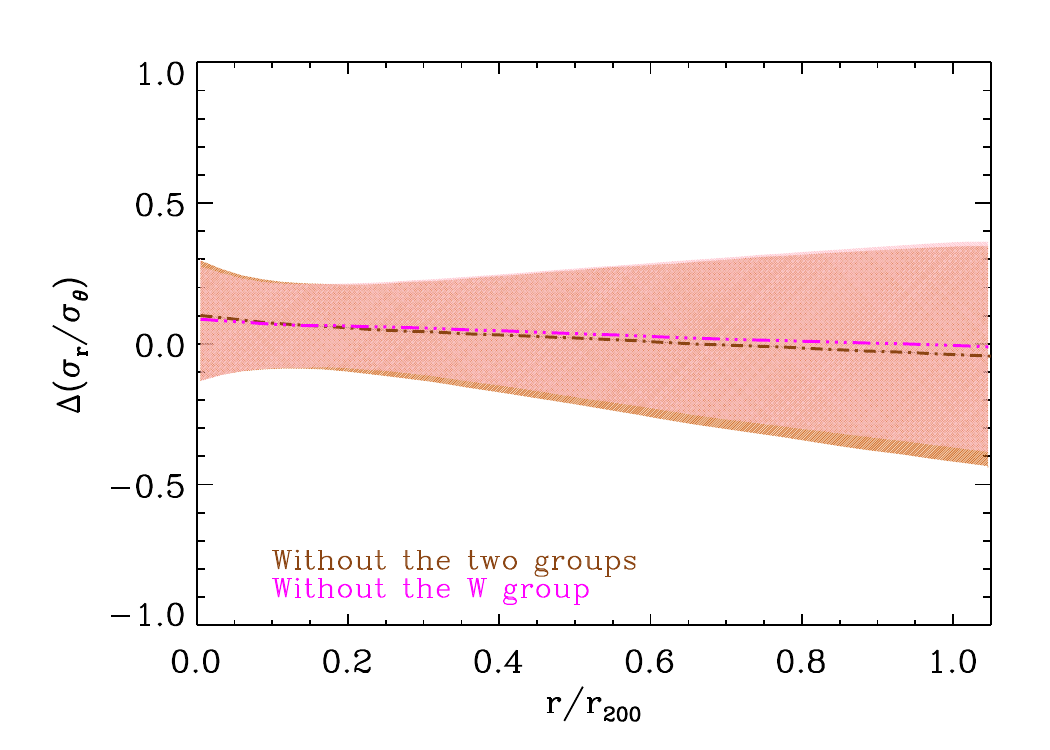}
    \caption{Brown dot-dashed line and shading (respectively: magenta triple-dot-dashed line and pink shading): difference between the velocity anisotropy profile obtained after removing the W and N groups (respectively: only the W group). The shading indicates the 1 $\sigma$ confidence interval of the difference.}
    \label{fig:beta_nosub}
\end{figure}

To assess the influence of these subclusters on the cluster dynamics, we redid the \mam\ analysis by removing the galaxies assigned to the W group, and, in a different run, all the galaxies assigned to the W and the N groups. The \mam\ results are very similar to, and not significantly different from, those obtained when all member galaxies without the exclusion of either one or both groups are considered. In Fig.~\ref{fig:beta_nosub} we show the differences between the velocity anisotropy profile obtained by removing either one or both groups and the velocity ansiotropy profile obtained using all cluster members. These differences are not significant, since they are well within the 1 $\sigma$ confidence intervals.
We conclude that even if A267 is not a fully relaxed cluster, its global dynamical state allows for a sensible application of the Jeans equation for dynamical equilibrium.

\subsection{The choice of the $M(r)$ and $\beta(r)$ models}
\label{sec:betadiffmodels}
Our results are based on the adoption of the
NFW model for $M(r)$ and the generalized Tiret model for $\beta(r)$ (see Sect. \ref{sec:method}). Here we consider relaxing these assumptions, by considering different models, namely the Burkert profile
\citep{Burkert1995} for $M(r)$ and the Osipkov-Merritt \citep{Osipkov1979,Merritt1985}
profile with two free parameters:
\begin{equation}
    \beta = \beta_0 + \beta_\infty r^2 / (r^2 + r^2_\beta),
    \label{eq:om}
\end{equation}
where $r_\beta=r_s$ as in the Tiret model (Eq.~\ref{eq:tiret}).

Using the Burkert profile \citep[a model that was found to best characterize systems with $\Delta m_{12} > 1.5$ systems in ][]{Zarattini2022}, the velocity anisotropy shifts to slightly more radial values at all radii compared to the solution obtained using the NFW profile, but the difference is not significant. When using the Osipkov-Merritt model, $\beta(r)$ is more radial near the centre and less radial in the outer region, which leads to a flattening of the profile compared to the Tiret case. However, in this case as well, the difference is marginal and not significant.

\subsection{Implications for FG formation and evolution models }
\label{sec:FG_model}

When FGs were proposed by \citet{Ponman1994}, they were thought to be the end point of galaxy group evolution, in which all the massive satellites had enough time to merge with the BCG, forming the dominant central galaxy together with the magnitude gap in a long-term project. In this model, FGs would be fossil relics of the ancient Universe, they were expected to be dynamically relaxed and with a small accretion rate, to preserve the magnitude gap. However, \citet{Sommer-Larsen2006} proposed that the infalling of galaxies on more radial orbits could be the mechanism responsible for the formation of the gap. This mechanism does not require a long and passive evolution, with no further accretions, since the timescale of merging between the BCG and the massive satellite is much faster when radial orbits are considered. This result was also supported by theoretical works, such as \citet{Kundert2017} and \citet{Kanagusuku2016}. By using the Illustris simulation, both papers found that the magnitude gap has a relatively recent formation and starts to appear at $z <0.3$. 

The results obtained in \citet{Zarattini2021} -- that galaxies falling into clusters with large magnitude gap ($\Delta m_{12} > 1.5$ in their work) are found on radial orbits, especially close to the virial radius -- perfectly fit in the \citet{Sommer-Larsen2006} model. However, in \citet{Zarattini2021} the analysis was based on the stacking of many different clusters and groups, and the magnitude gap was not large enough to prevent contamination from non-fossil systems, while in the present work we used a single, spectroscopically confirmed FG, with $\Delta m_{12} = 2.46$. Our cluster confirmed the results of the stacked clusters in \citet{Zarattini2021}, since we find the same shape for the anisotropy profile, with tangential orbits dominating the central region, where the BCG is found, and radial orbits continuously increasing towards the virial radius. 

Our results, therefore, re-enforce the \citet{Sommer-Larsen2006} idea that the accretion of galaxies on more radial orbits could be responsible for the formation of the magnitude gap. The next step in our understanding of FGs will be to find the link between radial orbits and the peculiar position of FGs within the large-scale structure.

\section{Conclusions}
\label{sec:conclusions}
In this work we analyse the velocity anisotropy profile of A267, a spectroscopically confirmed fossil cluster, with $\Delta m_{12} = 2.46$, which is part of the FOGO sample. To our knowledge, this is the first determination of the orbital shape of an individual FG. We find that the orbits of galaxies in the central cluster regions are mildly tangential, whereas in the external regions cluster galaxies move on radially elongated orbits. We found our result to be robust versus possible systematics, like the presence of subclusters and the use of different models for the mass and velocity anisotropy profiles.

Our result confirms our previous finding, in which a similar velocity anisotropy profile was found for a stack of $\Delta m_{12} > 1.5$ systems \citep{Zarattini2021}, and it is therefore supportive of the scenario proposed in \citet{Zarattini2023}. Central cluster galaxies in FGs sit at (or close to) the bottom of the potential well and move on tangential orbits, while galaxies in the external region, close to the virial radius, move on radially elongated orbits. These kind of orbits are conducive to stronger tidal forces as the infalling galaxies reach small pericentric radii, and these tidal forces can strip galaxies of part of their mass, or even destroy them. As a consequence, the central galaxy can accrete material from other cluster galaxies, which builds up the magnitude gap.

Moreover, according to \citep{Zarattini2023,Zarattini2022} FGs are found in peculiar positions in the cosmic web, far from nodes but close to filaments. It remains to be seen how this peculiar position of FGs in the large-scale structure of the Universe can be related to the orbital anisotropy of accreting galaxies. 

\begin{acknowledgements}

SZ acknowledges the financial support provided by the Governments of Spain and Arag\'on through their general budgets and the Fondo de Inversiones de Teruel, the Aragonese Government through the Research Group E16\_23R, and the Spanish Ministry of Science and Innovation and the European Union - NextGenerationEU through the Recovery and Resilience Facility project ICTS-MRR-2021-03-CEFCA. AB acknowledges the financial contribution from the INAF mini-grant 1.05.12.04.01 {\it "The dynamics of clusters of galaxies from the projected phase-space distribution of cluster galaxies"}. CPH acknowledges support from ANID through Fondecyt Regular project number 1252233. JALA acknowledge support from the Agencia Estatal de Investigaci\'on del Ministerio de Ciencia, Innovaci\'on y Universidades (MCIU/AEI) under grant WEAVE: EXPLORING THE COS- MIC ORIGINAL SYMPHONY, FROM STARS TO GALAXY CLUSTERS and the European Regional Development Fund (ERDF) with reference PID2023- 153342NB-I00 / 10.13039/501100011033.
\\
This article is partially based on observations made with the Nordic Optical Telescope (NOT) and Telescopio Nazionale Galileo (TNG) operated on the island of La Palma, in the Spanish Observatorio del Roque de los Muchachos of the Instituto de Astrof\'isica de Canarias. It is also partially based on date from the MMT Observatory, a joint facility of the Smithsonian Institution and the University of Arizona.
\\
Funding for the Sloan Digital Sky Survey IV has been provided by the Alfred P. Sloan Foundation, the U.S. Department of Energy Office of Science, and the Participating Institutions. SDSS acknowledges support and resources from the Center for High-Performance Computing at the University of Utah. The SDSS web site is www.sdss4.org. SDSS is managed by the Astrophysical Research Consortium for the Participating Institutions of the SDSS Collaboration including the Brazilian Participation Group, the Carnegie Institution for Science, Carnegie Mellon University, Center for Astrophysics Harvard \& Smithsonian (CfA), the Chilean Participation Group, the French Participation Group, Instituto de Astrof\'isica de Canarias, The Johns Hopkins University, Kavli Institute for the Physics and Mathematics of the Universe (IPMU) / University of Tokyo, the Korean Participation Group, Lawrence Berkeley National Laboratory, Leibniz Institut fur Astrophysik Potsdam (AIP), Max-Planck-Institut fur Astronomie (MPIA Heidelberg), Max-Planck-Institut fur Astrophysik (MPA Garching), Max-Planck-Institut fur Extraterrestrische Physik (MPE), National Astronomical Observatories of China, New Mexico State University, New York University, University of Notre Dame, Observatoio Nacional / MCTI, The Ohio State University, Pennsylvania State University, Shanghai Astronomical Observatory, United Kingdom Participation Group, Universidad Nacional Autonoma de Mexico, University of Arizona, University of Colorado Boulder, University of Oxford, University of Portsmouth, University of Utah, University of Virginia, University of Washington, University of Wisconsin, Vanderbilt University, and Yale University. 
\end{acknowledgements}
\bibliographystyle{aa}
\bibliography{bibliografia}
\end{document}